\documentclass[12pt]{article}
\def\be{\begin{equation}}
\def\ee{\end{equation}}
\def\bc{\begin{center}}
\def\ec{\end{center}}
\def\bea{\begin{eqnarray}}
\def\eea{\end{eqnarray}}
\def\dd{\displaystyle}
\def\nn{\nonumber}

\def\crbig{\\\noalign{\vspace {3mm}}}

\def\Re{{\rm Re\,}}
\def\Im{{\rm Im\,}}
\def\ov{\overline}

\def\cL{{\cal L}}

\catcode`@=11
\def\marginnote#1{}
\newcount\hour
\newcount\minute
\newtoks\amorpm
\hour=\time\divide\hour by60
\minute=\time{\multiply\hour by60 \global\advance\minute by-\hour}
\edef\standardtime{{\ifnum\hour<12 \global\amorpm={am}%
        \else\global\amorpm={pm}\advance\hour by-12 \fi
        \ifnum\hour=0 \hour=12 \fi
        \number\hour:\ifnum\minute<10 0\fi\number\minute\the\amorpm}}
\edef\militarytime{\number\hour:\ifnum\minute<10 0\fi\number\minute}
\def\draftlabel#1{{\@bsphack\if@filesw {\let\thepage\relax
   \xdef\@gtempa{\write\@auxout{\string
      \newlabel{#1}{{\@currentlabel}{\thepage}}}}}\@gtempa
   \if@nobreak \ifvmode\nobreak\fi\fi\fi\@esphack}
        \gdef\@eqnlabel{#1}}
\def\@eqnlabel{}
\def\@vacuum{}
\def\draftmarginnote#1{\marginpar{\raggedright\scriptsize\tt#1}}
\def\draft{\oddsidemargin 0.0truein
        \def\@oddfoot{\sl preliminary draft \hfil
        \rm\thepage\hfil\sl\today\quad\militarytime}
        \let\@evenfoot\@oddfoot \overfullrule 3pt
        \let\label=\draftlabel
        \let\marginnote=\draftmarginnote
   \def\@eqnnum{(\theequation)\rlap{\kern\marginparsep\tt\@eqnlabel}%
\global\let\@eqnlabel\@vacuum}  }
\catcode`@=12
\begin{document}
\begin{titlepage}
\vspace*{-1cm}
\phantom{hep-th/0403043} 
\hfill{NEIP-04/02}
\\
\phantom{hep-th/0403043} 
\hfill{LPTENS-04/08}
\\
\phantom{hep-th/0403043}
\hfill{CERN-PH-TH/2004-040}
\vskip 2.0cm
\bc
{\Large \bf Potentials and superpotentials in the effective}
\ec
\bc
{\Large\bf \boldmath{$N=1$} supergravities from higher dimensions}
\ec
\vskip 1.0  cm
\begin{center}
{\large Jean-Pierre Derendinger}~\footnote{e-mail address: 
jean-pierre.derendinger@unine.ch}~,
\\
\vskip .1cm
Institut de Physique, Universit\'e de Neuch\^atel
\\
CH-2000  Neuch\^atel, Switzerland
\\
\vskip .2cm
{\large Costas Kounnas}~\footnote{e-mail address: 
costas.kounnas@lpt.ens.fr}~,
\vskip .1cm 
Laboratoire de Physique Th\'eorique, Ecole Normale Sup\'erieure
\\
24 rue Lhomond, F-75231 Paris Cedex 05, France
\\
\vskip .1cm
and
\\
\vskip .2cm
{\large Fabio Zwirner}~\footnote{e-mail address:
fabio.zwirner@roma1.infn.it}$^{\, ,}$~\footnote{On leave from
Universit\`a di Roma `La Sapienza' and INFN, Sezione di Roma, Italy}
\\
\vskip .1cm
CERN, Physics Department, Theory Division,
\\ 
CH-1211 Geneva 23, Switzerland
\end{center}
\vskip 1.0cm
\begin{abstract}
\noindent
We consider $N=1$ superpotentials corresponding to gaugings of an
underlying extended supergravity for a chiral multiplet in the
$SU(1,1)/U(1)$ manifold of curvature $2/3$. We analyze the resulting
$D=4$ scalar potentials, and show that they can describe different
$N=1$ phases of higher-dimensional supergravities, with broken or
unbroken supersymmetry, flat or curved backgrounds, sliding or
stabilized radius. As an application, we discuss the $D=4$ effective
theory of the detuned supersymmetric Randall-Sundrum model in two
different approximation schemes.
\end{abstract}
\end{titlepage}
\setcounter{footnote}{0}
\vskip2truecm
\setlength{\baselineskip}{.7cm}
\bc
{\bf 1.~Introduction and summary}
\ec
Compactifications of higher-dimensional supergravity and superstring
theories preserving $N=1$ supersymmetry in $D=4$ dimensions, in its exact or spontaneously broken phase, have
great phenomenological interest. Their low-energy effective theories
\cite{effn1} typically include a chiral multiplet whose spin-zero
component has two real degrees of freedom: one parameterizes the
volume of the internal space, the other one some internal gauge degree
of freedom of the higher-dimensional theory. If we neglect the
dynamics of all other fields, apart from the gravitational
multiplet, this complex scalar field parameterizes the special
K\"ahler manifold $SU(1,1) / U(1)$, with scalar curvature $2/3$. In a
suitable field basis, we can decompose it as
\be
\label{Tdef}
T = t + i \, \tau \, ,
\qquad \qquad
(t  \,\, {\rm and}\,\,  \tau \,\,{\rm real}) \, ,
\ee
and write for its K\"ahler potential, in the standard notation of
$N=1$, $D=4$ supergravity~\cite{cfgvp} and in $D=4$ Planck mass units:
\be
\label{KahlerT}
K(T,\ov{T}) = - 3 \, \log (T+\ov T) \, .
\ee
This is the case, for example, for the known compactifications of
minimal $D=5$ supergravity \cite{pure5d,gauged5d} on the orbifold
$S^1/Z_2$, both in the flat \cite{bfzjhep} and in the warped
\cite{rsefftun,rseffdetun} case. This is also true for
compactifications of the same theory on the circle $S^1$ \cite{gauged5d}, which give
$N=2$, $D=4$ supergravity coupled to a single vector multiplet. The
same result is obtained in those higher-dimensional supergravity and
superstring compactifications with branes, orientifolds and fluxes,
where some of the moduli are fixed, but not the overall volume modulus
\cite{fluxes}.

In agreement with the special geometry of $N=2$, $D=4$ supergravity
coupled to vector multiplets \cite{dwvp,seven}, the K\"ahler potential
(\ref{KahlerT}) derives from the prepotential:
\be
\label{prep}
F = {(X^1)^3 \over X^0} \, .
\ee
Indeed \cite{seven,dfmp} :
\be
K = - \log Y \, ,
\;\;\;\;\;
Y = i \left( \ov{X}^I F_I - X^I \ov{F}_I \right) \, ,
\label{kahler2}
\ee
where a sum over the index $I=0,1$ is understood, $F_I \equiv \partial
F / \partial X^I$, and $X^0$ is the compensating vector multiplet of
the $N=2$ superconformal theory. Introducing the unconstrained field
$T = i X^1 / X^0$, and choosing the gauge $X^0 = 1$, we obtain $Y = (
T + \ov{T} )^3$, which gives precisely the K\"ahler potential
(\ref{KahlerT}).

Given the K\"ahler potential (\ref{KahlerT}), the superpotential
$w(T)$ encodes the information on the underlying higher-dimensional
model. Schematically, considering only the $T$ field and a generic
$w(T)$ amounts to reduce supersymmetry to four supercharges. The
structure of $w(T)$, however, should keep track of its
higher-dimensional origin. With sixteen supercharges, as required minimally by the
ten-dimensional supersymmetry algebra, a four-dimensional supergravity is completely 
specified by
the real structure constants defining the gauging. Truncating this
$N=4$ theory \cite{N=4} to $N=1$ leads to seven moduli (three associated with the
complex structure, $T_1, T_2, T_3$, three associated with the K\"ahler structure,
$U_1, U_2, U_3$, and the four-dimensional dilaton) or more. The $N=1$ 
superpotential can then be obtained directly from the field-dependent 
gravitino mass term\footnote{The procedure outlined here is described in 
detail in ref. \cite{adk}, section 4.}
\be
m_{3/2} = 
w \, e^{K/2} = (S + \ov{S})^{-1/2} \left( f_{i_1 i_2 i_3} \, 
\Phi^{i_1} \, \Phi^{i_2} \, \Phi^{i_3} + S \, 
\widetilde{f}_{i_1 i_2 i_3} \, \Phi^{i_1} \, \Phi^{i_2} 
\, \Phi^{i_3} \right) \, ,
\label{n4gau}
\ee
where $S$ is the complex dilaton parameterizing the $SU(1,1)/U(1)$
manifold. The $\Phi^{i_A}$ collectively denote all the other scalar
fields of the $N=1$ truncation, with $A=1,2,3$ labeling the threefold
degeneracy of the spectrum (the three $T_A$ and $U_A$ moduli for 
instance)\footnote{The exact relation between $\Phi^{i_A}$ and, for instance,
$T_A$ and $U_A$ follows from the solution of the $N=4$ Poincar\'e
constraints.}, 
and $(f_{i_1 i_2 i_3}, \, \widetilde{f}_{i_1 i_2 i_3})$
are the real structure constants defining the gauging of the
underlying $N=4$ theory. These structure constants also induce, in
general, a scalar potential.  Notice that the expression (\ref{n4gau})
contains both perturbative and non-perturbative terms (with respect to
$S$), as dictated by the $SU(1,1)$ duality of $N=4$ supergravity. The
procedure outlined above is general enough to describe the dynamics of
a variety of phases of the higher-dimensional theory in a
duality-invariant way. Examples include effective Lagrangians for $N=4$ strings 
\cite{GivP}, finite-temperature phases of
five-dimensional superstrings \cite{adk, ak} and the effective description of partial
(non-)perturbative supersymmetry breaking in string theory \cite{kk}. Freezing all the scalar
fields $\Phi^{i_A}$ apart from $T$ will certainly lead to a polynomial
superpotential $w(T)$ of third order in $T$. Its coefficients will be
defined by the values assigned to the frozen fields, which are in
general complex. Actually, it has been known for long \cite{hetflux}
that non-zero coefficients can be related to background values of
antisymmetric tensors, {\em aka} fluxes. We will therefore consider a
superpotential of the form
\be
w (T) = m_0 - i \, m_1 \, T + 
3 \, n^1 \, T^2 + i \, n^0 \, T^3  \, ,
\label{sup2}
\ee
where $(m_0,m_1,n^0,n^1)$ are four arbitrary complex coefficients.
Notice also that, since
\be
w = m_I \, X^I - n^I \, F_I \, ,
\label{sup1}
\ee
a generic $N=2$ gauging of the theory \cite{fklz} with one vector multiplet only
would correspond to the additional condition of taking the
coefficients $(m_0,m_1,n^0,n^1)$ to be real, apart from an invisible
overall phase in $w$. Known examples of $N=2$ gaugings with one vector multiplet
correspond to
supersymmetric $AdS_5$ backgrounds \cite{gauged5d}, and to the $D=4$
effective theories of supersymmetry-breaking compactifications on a
flat $D=5$ background \cite{fkpz, PZ}, induced by twisted periodicity
conditions {\em \`a la} Scherk-Schwarz \cite{scsc}. Some orientifold
compactifications provide instead examples of superpotentials of the
form (\ref{sup2}), but with complex coefficients \cite{fluxes}. We
will discuss in this paper another instructive example of this sort: 
the effective theory of the supersymmetric Randall-Sundrum (SUSY-RS) model
with two branes \cite{rs1} and arbitrary tensions \cite{susyrstun,
rsefftun, kaloper, susyrsdetun, rseffdetun}.

Our goal in this paper is to study the general features of the $T$
gaugings, in the generalized sense defined by eq.~(\ref{sup2}) with
complex coefficients. After giving a complete discussion of the `true'
$N=2$ gaugings, we confront the superpotential (\ref{sup2}) with
another requirement, motivated by $N=1$ compactifications of
higher-dimensional supergravities with negligible warping: the
independence of the scalar potential from $\tau$, which in this
context is proportional to the internal component of an abelian gauge
field. We then perform a similar analysis in a field basis that is
more appropriate to discuss warped compactifications with
non-negligible warping. We conclude the paper by discussing, as an
illustration of our formalism, the effective theory of the detuned
SUSY-RS model \cite{rsefftun,rseffdetun} in two different
approximations. Some useful formulae for the backgrounds
\cite{kaloper} of the SUSY-RS model are collected in the appendix.
The generalization of the results of this paper to the case of more
chiral multiplets, relevant for the effective theories of superstring
compactifications with fluxes, branes and orientifolds, is left for
future work.
\vspace{0.5cm}
\bc 
{\bf 2. `True' \boldmath{$T$}--gaugings} 
\ec
We recall first that the scalar potential of a model with the K\"ahler
potential of eq.~(\ref{KahlerT}) and a generic superpotential $w(T)$
is
\be
V = {1 \over ( T + \ov{T} )^2 } \left[ {|w_T|^2  ( T + \ov{T} ) 
\over 3} - (w_T \ov{w} + \ov{w}_{\ov{T}} w ) \right]  ,
\label{VTgen}
\ee
and also that the K\"ahler potential of eq.~(\ref{KahlerT}) has an
$SU(1,1)$ K\"ahler invariance, identified as a continuous
$T$--duality:
\be
T \quad \longrightarrow \quad {aT-ib \over icT+d} \, , 
\;\;\; \qquad 
a,b,c,d \,\ {\rm real} ,
\qquad 
ad-bc=1 .
\label{duality}
\ee
Invariance of the theory requires the following transformation of the
superpotential:
\be
w(T) \quad \rightarrow \quad (icT+d)^3 \, w \left(
{aT-ib \over icT+d} \right) ,
\label{wdual}
\ee
which turns a cubic polynomial into another one, with transformed
parameters. $T$--duality can then be used to eliminate some of them,
as discussed below. Notice that both $SU(1,1)$ duality and
this form-invariance of the superpotential find their origin in the 
underlying $N=4$ duality symmetry. 

We concentrate in this section on `true' $N=2$ gaugings where the
superpotential parameters $(m_0,m_1,n^0,n^1)$ in eq.~(\ref{sup2}) are
real numbers. With the help of eq.~(\ref{wdual}), we can show that
$SU(1,1)$ can always be used to eliminate the cubic and linear term in 
$w(T)$, i.e. to set $n^0 = m_1 = 0$ (notice that, even if we start
from real parameters and we allow for duality transformations, the
general form of the superpotential includes at least an ``electric''
and a ``magnetic'' term). The proof is more easily given using the
generators
\be
\begin{array}{rrcll}
{\cal I}:& \qquad T&\quad\longrightarrow\quad & 1/T \, , \qquad & 
(a=d=0, \,\, b=-c=1), \crbig
{\cal S}:& \qquad T&\quad\longrightarrow\quad & T-i \, , & 
(a=b=d=1,\,\, c=0). 
\end{array}
\ee 
Under inversion ${\cal I}$, the superpotential parameters transform
as~\footnote{Since $b=-c$, ${\cal I}$ squares to $-1$ when acting on
the superpotential.}
\be
(m_0,m_1, 3n^1, n^0) \quad\longrightarrow\quad (-n^0, -3n^1, m_1, 
m_0).
\ee
A generic element of $SU(1,1)$ is then of the form ${\cal S}^C{\cal
I}{\cal S}^B{\cal I}{\cal S}^A$, ($A,B,C$ real). First, by performing 
an appropriate duality transformation, we can
always eliminate the cubic term, {\it i.e.} set $n^0=0$: it is sufficient to
consider a shift of $T$ to eliminate the constant term, followed by an
inversion. If the resulting superpotential is a quadratic polynomial,
a shift of $T$ eliminates then the linear term. If it is linear in $T$, a shift 
of $T$ is used to eliminate the constant term and an inversion leads then
to a single quadratic term.  However, the
discussion will be simple enough if we only eliminate the $T^3$ term,
$n^0=0$, leaving $m_0$, $m_1$ and $n^1$ free.

With $n^0=0$, the scalar potential has the simple form
\be
V = - {1 \over 3 t} \left[ m_1^2 + 9 \, m_0 \, n^1 - 3 \, m_1 \,
n^1 \, \tau + 9 \, (n^1)^2 (t^2 + \tau^2) \right] \, .  
\label{simplepot}
\ee 
With $n^0 \ne 0$, the potential depends on the same powers of $t$ and
$\tau$, but with more complicated numerical coefficients. Relaxing the
requirement of real coefficients in the superpotential does not allow
to remove the cubic term by duality transformations, and leads to
additional powers of $t$ and $\tau$ in the scalar potential.

To discuss the phases of the theory, we need the expression of $f^T
\equiv (2t)^{-1/2} [2 \, w_T \, t / 3 - w]$, the auxiliary field which
controls supersymmetry breaking, as well as the field-de\-pen\-dent
gravitino mass:
\be
\label{m3/2T}
m_{3/2}^2 = |w|^2 \, e^K = -{V \over 3} + (2t)^{-2}|f^T|^2 \, .
\ee
The relation $V = - 3 \, m_{3/2}^2$ holds then for a supersymmetric
$AdS_4$ phase.

Taking into account that the allowed field configurations correspond
to $t>0$ and arbitrary $\tau$, we can now study the properties of
eqs.~(\ref{simplepot})--(\ref{m3/2T}) for different values of the
gauging parameters and the corresponding allowed phases of the theory.
\bc
(I): $n^1=m_1=0$, $m_0 \ne 0$.
\ec
This case corresponds to the original no-scale model \cite{noscale}
with $V \equiv 0$, and spontaneously broken $D=4$ supersymmetry in
Minkowski space-time, with $m_{3/2}^2 = (2t)^{-3} m_0^2$. This
includes the effective theory of Scherk-Schwarz compactifications
\cite{scsc} of pure ungauged $D=5$ supergravity on the orbifold
$S^1/Z_2$, with a flat and constant bosonic background
\cite{fkpz,bfzjhep}. At the full $N=2$ level, this case would
correspond to the $D=4$ effective theory of a Scherk-Schwarz
compactification on the circle $S^1$, of which the $N=1$ theory is a
consistent $Z_2$ truncation. Notice finally that a generic duality
transformation (\ref{wdual}) leads to the equivalent class of
superpotentials:
\be
w(T) = m_0 \, (i c T + d)^3 \, ,
\label{gennoscale}
\ee
representing the most general no-scale model in the $T$ field basis.
\bc
(II): $n^1=m_0=0$, $m_1 \ne 0$.
\ec
In this case the potential is negative definite, does not depend on
$\tau$ and does not have any stationary point with respect to $t$:
$V=-m_1^2/(3 t)$. The educated reader will immediately notice that,
since $m_{3/2}^2=m_1^2(t^2+ \tau^2)/(8t^3)$, for $\tau = 0$ we get the
$AdS_5$ relation $V = - (8/3) \, m_{3/2}^2$. This is because, in the
context of pure $D=5$ supergravity, $m_1$ is proportional to the
gauging parameter of the graviphoton. This gauging leads to $AdS_5$
supergravity~\cite{gauged5d}. The relation with the present case can
be established, for instance, by writing the $D=5$ gravitino variation
on a background with $D=4$ Minkowski symmetry, as in the bulk of the
SUSY-RS model. The result is identical to the gravitino variation with
a superpotential linear in $T$. This is nothing more than a formal
manipulation unless a boundary is applied to $AdS_5$, as in the
$S^1/Z_2$ orbifold: in this case, however, boundary contributions do
modify the effective $D=4$ superpotential.  The equivalent case
\bc
(III): $m_0, \, m_1 \ne 0$, $n^1 = 0$.
\ec
is generated by the shift $T \rightarrow T + i m_0 / m_1$.
\bc
(IV): $m_0=m_1=0$, $n^1 \ne 0$.
\ec
This is the case of a purely ``magnetic'' gauging. It is equivalent to
the previous case since it is connected to it by an inversion ${\cal
I}$: the potential depends then only on the real part of $(1/T)$.
\bc
(V): $m_0, \, n^1 \ne 0$.
\ec
In the case where both electric and magnetic terms are present, it is
not restrictive to set $m_1=0$, since it can always be reached by a
suitable duality transformation. As a function of $\tau$, the
potential has a local maximum at $\tau = 0$. It then has a stationary
point, at
\be
t^2 = {m_0 \over n^1}  \, ,
\qquad \qquad
\tau = 0 \, ,
\ee
if and only if $m_0 n^1 > 0$. At this point,
\be 
V = - 6 (n^1)^2 t = - 3 m_{3/2}^2 \, , \qquad \qquad f^T = 0 \, ,
\ee 
and one has a stable supersymmetric $AdS_4$ phase \cite{stability}.
If, however, $m_0 n^1 < 0$, the field $t$ is not stabilized.

To summarize, `true' $N=2$ $T$--gaugings offer the following
possibilities: broken supersymmetry in flat space, with a complex flat
direction; unstable potential with one or no axionic flat direction,
identified with $\tau$ up to a duality transformation; unbroken
supersymmetry in $AdS_4$ with stabilized complex $T$. As will be
discussed in detail later, none of these solutions is appropriate to
describe the detuned SUSY-RS model.
%
\newpage
\bc 
{\bf 3. Generalized \boldmath{$T$}--gaugings} 
\ec
In the general case of the superpotential (\ref{sup2}) with complex
coefficients, we are interested in finding all cases where the scalar
potential is independent of the axion $\tau$. This is suggested by the
fact that, in compactifications with negligible warping, $\tau$ is
proportional to the internal component of an abelian gauge field.  The
general solution to the above problem allows for a unique possibility,
besides those corresponding to `true' $N=2$ gaugings discussed in
section~2. Absorbing as usual an overall phase:
\be
\label{supa}
{\rm (VI):}\,\,  w (T)  = \rho_0 + \rho_1 \, e^{\dd i \, \varphi} \, T \, ,
\qquad
(\rho_0, \rho_1, \varphi \,\,{\rm real} \, ,
\; \rho_0 , \rho_1 \ge 0) \, . 
\ee
This case corresponds to $m_0 = \rho_0$, $m_1 = \rho_1 e^{i ( \varphi
+ \pi/2 )}$, $n^0=n^1=0$, and is a generalization of cases (I)-(III)
discussed in section~2. We then concentrate on the novel possibility
represented by $\rho_0, \rho_1, \cos \varphi \ne 0$: as we will see in
section~5, this corresponds to the effective theory of the detuned
SUSY-RS model in the limit of small warping, when the compactification
radius is small with respect to the $AdS_5$ radius. The scalar
potential is:
\be
V = - {\rho_1 \over 6 \, t^2}
( 3 \, \rho_0 \, \cos \varphi + 2 \, \rho_1 \, t ) \, ,
\label{pota}
\ee
and has a minimum for $t=-3 \, \cos \varphi \, \rho_0 / \rho_1$, which
falls within the allowed field configurations if $\cos \varphi <
0$. At the minimum $V=\rho_1^3/(18 \, \rho_0 \cos \varphi) < 0$,
corresponding to an $AdS_4$ background. Supersymmetry is unbroken for
$\tau = \rho_0 \, \sin \varphi / \rho_1$, otherwise it is
spontaneously broken. This is an example of $t$ stabilization in
$AdS_4$.

To conclude this section, we notice that the decomposition of
eq.~(\ref{Tdef}) is stable under imaginary shifts, but unstable
under the inversion of $T$, which mixes its real and imaginary
part. Therefore, the requirement of a $\tau$-independent potential
translates into the existence of a real flat direction when acting
with the full $SU(1,1)$ duality group. Similar considerations apply to
general analytic field redefinitions.
\vspace{0.5cm}
\bc 
{\bf 4. Potentials and superpotentials in the SUSY-RS basis} 
\ec
Another parameterization of the $SU(1,1)/U(1)$ K\"ahler manifold of
curvature $2/3$, equivalent to the one considered in the previous
sections, is
\be
\label{Kbr}
K (U,\ov{U}) = - \log \, Y  ,
\qquad \qquad
Y = \pm \left[ e^{\dd \lambda \pi (U + \ov{U})} - 1 \right]^3  .
\ee
This particular parameterization has been used for writing the
effective theory of the SUSY-RS model~\cite{rsefftun, rseffdetun}, in
the limit of small $D=4$ cosmological constant. Indeed, our notation
in eq.~(\ref{Kbr}) has been chosen to fit such an interpretation, even
if the present discussion has a more general validity. As explained in
section~5 and in the appendix, we can identify $\lambda$ with the mass
scale of the $D=5$ cosmological constant, in units of the $D=5$ Planck
mass. Moreover, we can set
\be
\label{udef}
U= r + i \, b \, , 
\ee
where the `radion' $r(x)$ describes the $D=4$ scalar fluctuation of
the $D=5$ metric, and the `axion' $b(x)$ is proportional to the zero
mode of the internal component $B_5$ of the graviphoton. Because of
the abelian $D=5$ gauge invariance, the scalar potential cannot depend on
$b(x)$.

We can now introduce, for notational convenience, the auxiliary variables
\be
X = e^{\dd \lambda \pi U} = e^{\dd \lambda \pi (r + i b)} ,
\qquad
\ov{X} = e^{\dd \lambda \pi \ov{U}} = e^{\dd 
\lambda \pi (r - i b)} .
\label{XUdef}
\ee
The relation between the parameterizations of eqs.~(\ref{Kbr}) and
(\ref{KahlerT}) is then given by the analytic field redefinition
\be
\label{TXXT}
T(X) = \pm {X-1\over X+1} \, , 
\qquad 
X(T) = {1 \pm T \over 1 \mp T } \, ,
\ee
which induces, neglecting as usual an overall phase, the following
transformation of the superpotential:
\be
\label{Wbr}
w(X) = w[T(X)] {(1+X)^3 \over2\sqrt2}  \, ,
\qquad
w(T) = w[X(T)] {(1 \mp T)^3\over2\sqrt2}  \, .
\ee  
To reach the conventions of refs.~\cite{rsefftun,rseffdetun}, we 
would need an additional K\"ahler transformation:
\be
\label{Ktransf}
w(X) \rightarrow w(X) X^{-3} \, ,
\qquad
(X \ov{X} - 1) \rightarrow (X \ov{X} - 1) (X \ov{X})^{-1}
= [1 -  (X \ov{X})^{-1}] \, .
\ee
The superpotential $w(X)$, corresponding to the superpotential $w(T)$
in (\ref{sup2}) via (\ref{Wbr}), is again a cubic polynomial. As in
the previous section, we drop here the requirement of real
coefficients.

For a generic superpotential $w[X(U)]$, the corresponding
scalar potential reads
\be
V( U , \ov{U} )= \pm {|w_U|^2 [ 1 - e^{- \lambda \pi (U + \ov{U})} ] 
- 3 \lambda \pi (w_U \ov{w} + \ov{w}_{\ov{U}} w ) + (3 \lambda 
\pi)^2 |w|^2 \over 3 \lambda^2 \pi^2 [e^{\lambda \pi (U + \ov{U})}
- 1]^2} \, ,
\label{VgenU}
\ee
or, equivalently,
\be
V( X , \ov{X} )= \pm {|w_X|^2 ( X \ov{X} - 1 )  
- 3 (X \, w_X \, \ov{w} + \ov{X} \, \ov{w}_{\ov{X}} \, w ) 
+ 9 |w|^2 \over 3 ( X \ov{X} - 1)^2 } \, ,
\label{VgenX}
\ee
where the sign ambiguity comes from the definition of the $Y$
function, eq.~(\ref{Kbr}). If we now ask for a non-trivial
superpotential $w \ne 0$ such that the $D=4$ scalar potential does not
depend on the imaginary part of $U$ (the phase of $X$), as required by
$D=5$ gauge invariance, and we factor out for convenience an arbitrary
phase, we obtain as general solution the four possibilities listed
below.
\be
{\rm (i):} \,\, w(X) = \rho (1 + e^{\dd i \varphi} \, X)^3 \, ,
\qquad
(\rho, \varphi \,\, {\rm real }, \; \rho > 0) \, .
\label{sup1x}
\ee
This leads to a no-scale model with identically vanishing scalar
potential, $V \equiv 0$. Both signs in eq.~(\ref{Kbr}) are allowed, if
we restrict the allowed field configurations to $|X| > 1$ and $|X| <
1$, respectively. Supersymmetry is spontaneously broken for all
allowed values of $X$.
\be
\label{sup2x}
{\rm (ii):} \,\,  w(X) = \rho_0 + \rho_3 \,  e^{\dd i \varphi} \, X^3 \, ,
\qquad
(\rho_0, \rho_3, \varphi \,\, {\rm real},
\; \rho_0, \rho_3 \ge 0) \, .
\ee
This leads to the scalar potential
\be
V = \pm {3 [\rho_0^2 - \rho_3^2 (X \ov{X})^2] \over
(X \ov{X} - 1)^2} \, .
\label{potx2}
\ee
This case includes the effective theory of the detuned Randall-Sundrum
model in the limit of small $D=4$ cosmological constant, as derived in
\cite{rseffdetun} and discussed in section~5. The potential has
stationary points for $X=0$ and, if $\rho_3 > 0$, for
$|X|=\rho_0/\rho_3$. The case $X=0$ is acceptable only if we choose
the minus sign in eqs.~(\ref{Kbr}) and (\ref{potx2}): it corresponds
to a stable vacuum with unbroken supersymmetry, $m_{3/2}^2 = - \langle
V \rangle / 3 = \rho_0^2$; for $\rho_0 = 0$ it is Minkowski, and there
is a classically massless complex scalar, even if with $\rho_3 > 0$
the potential has a quartic term with positive coefficient for the
radion; for $\rho_0 > 0$ it is $AdS_4$. To discuss the other
stationary point we must consider $|X| < 1$ and the minus sign in
eqs.~(\ref{Kbr}) and (\ref{potx2}) if $\rho_0 < \rho_3$, $|X|>1$ and
the plus sign if $\rho_0 > \rho_3$ (for $\rho_0 = \rho_3$ there is a
singularity).  Both cases give the same physics. At the minimum the
potential is negative, $\langle V \rangle = - 3 \, \rho_0^2 \,
\rho_3^2 \, / \, | \rho_3^2 - \rho_0^2 | < 0$, but its second
derivative with respect to $|X|$ is positive, so we have a stable
$AdS_4$ vacuum. Supersymmetry is spontaneously broken unless $3 \,
\lambda \, \pi \, b + \varphi = \pm \pi$.
\be
{\rm (iii):} \,\, w(X) = \rho \, X^2 \, ,
\qquad
(0 < \rho \,\, {\rm real}) \, .
\ee
The scalar potential reads
\be
V = \pm {\rho^2 (X \ov{X})(X\ov{X}-4) \over
3 (X \ov{X} - 1)^2} \, .
\label{potx3}
\ee
Its only stationary point is for $X = 0$, so we must choose the minus
sign in eqs.~(\ref{Kbr}) and (\ref{potx3}), which allows for the field
configurations with $|X| < 1$. This leads to unbroken supersymmetry in
a flat background, with a stable minimum at $\langle X \rangle =
0$. If this is interpreted as the effective theory coming from the
compactification of a higher-dimensional supergravity, it amounts to
the simplest example of moduli stabilization with unbroken
supersymmetry in flat space.  Notice that a similar result could be
obtained in a much larger class of theories, characterized by: a
K\"ahler potential $K(|X|^2)$, with arbitrary functional form as long
as it does not depend on the phase of $X$ and it admits $X=0$ among
the allowed field configurations; a monomial superpotential $w(X)
\propto X^n$, with $n \ge 2$.
\be
{\rm (iv):} \,\, w(X) = \rho \, X \, ,
\qquad
(0 < \rho \,\, {\rm real}) \, .
\ee
The scalar potential reads
\be
V = \pm { \rho^2 ( 4 X \ov{X} - 1 ) \over
3 ( X \ov{X} - 1 )^2 } \, .
\ee
Its only stationary point is for $X = 0$, so we must choose the minus
sign in eqs.~(\ref{Kbr}) and (\ref{potx3}), which allows for the field
configurations with $|X| < 1$. However, it can be easily checked that
in this case $X=0$ is an unstable $dS_4$ maximum.
\vspace{0.5cm}
\bc 
{\bf 5. The effective theory of the detuned SUSY-RS model} 
\ec
As an illustration of the formalism described in the previous
sections, we conclude by discussing the effective theory of
the SUSY-RS model, in its detuned version. We shall see that such an
effective theory can be formulated both in the $T$ basis and in the SUSY-RS
basis, in two different approximations. In both cases, it corresponds
to a specific example of the generalized gaugings classified in
sections~3 and 4.

The effective $N=1$, $D=4$ theory of the SUSY-RS model is already
known, both in the tuned \cite{rsefftun} and in the detuned
\cite{rseffdetun} case. We will present here an alternative,
technically simpler derivation of its K\"ahler potential and
superpotential, both in the SUSY-RS field basis of section~4 and in the $T$
field basis of section~3. For the reader's convenience, our notation
and some useful results on the backgrounds of the SUSY-RS model are
spelled out in the appendix. With a slight variation with respect to
\cite{rsefftun, rseffdetun}, we define the radion field $r(x)$ in such
a way that its VEV coincides with $r_c$, the compactification radius in units
of the $D=5$ Planck mass $M_5$. This will allow us to derive
the effective theory for the radion by simply replacing $r_c$ with
$r(x)$ in the background ansatz:
\be
\label{inttext}
ds^2 = a^2 (r_c,y) \, \hat{g}_{\mu \nu} (x) 
dx^\mu dx^\nu + r_c^2 \, dy^2 \, ,
\ee
where the explicit form of $a^2 (r_c,y)$ and the details of the
notation are given in the appendix. This replacement should not be
done, of course, in the equations that relate $r_c$ with the input
parameters $(\lambda, \, \lambda_0, \, \lambda_\pi)$. Because of
supersymmetry we can consider only the bosonic part of the $D=5$
supergravity Lagrangian. Moreover, since we know from the tadpole
analysis of \cite{rsefftun, rseffdetun} the correct complexification
of the radion and axion degrees of freedom, we can consider only the
gravitational part ${\cal L}_g$ of the $D=5$ bosonic Lagrangian, whose
explicit form is given in the appendix, and neglect the part
containing the graviphoton. After the replacements
\be
r_c \longrightarrow r(x) ,
\qquad
a(r_c,y) \longrightarrow a[r(x),y]  ,
\label{replac}
\ee
we obtain, neglecting as usual total derivatives:
\bea
\cL_g & = & 
- {1 \over 2} \Phi \widehat{e}_4 \widehat{R} + 
{3 \over 4} \Phi \widehat{e}_4 \widehat{g}^{\mu \nu}
(\partial_\mu \log \Phi)(\partial_\nu \log \Phi) -
{3 \over 4} \Phi \widehat{e}_4 \widehat{g}^{\mu \nu}
(\partial_\mu \log r)(\partial_\nu \log r)
\nn \\ & & 
+ 6 \, \widehat{e}_4 \,
\left[ { (a')^2 \, a^2 \over r}
+ \lambda^2 \, r \, a^4
- \lambda_0 \, a^4 \, \delta(y)
+ \lambda_\pi \, a^4 \, \delta(y - \pi) 
\right] .
\label{5Defflag}
\eea
Notice that the $D=4$ Einstein term has an $x$-and-$y$-dependent $D=5$
dilaton prefactor:
\be
\label{bigphiofx}
\Phi [r(x), y] = r(x) \, a^2  [r(x), y] \, .
\ee

We begin by observing that the radion kinetic term in
eq.~(\ref{5Defflag}) is compatible with a $y$-dependent K\"ahler
potential
\be
\label{ku}
K = - 3 \log (U + \ov{U}), 
\qquad\qquad 
\Re U = r(x) \, .
\ee
This is confirmed by the axion kinetic term, which arises from the
$D=5$ Maxwell term for the graviphoton $B_M$. Putting $B_5=\sqrt{3/2}
\, b(x)$, we get:
\be
- {1 \over 4} B_{MN} B^{MN} = - {3\over 4} \widehat{e}_4 \Phi 
G^{55} \widehat{g}^{\mu\nu} (\partial_\mu b)(\partial_\nu b) + 
\ldots = - {3\over4} \widehat{e}_4 \Phi r^{-2} \widehat{g}^{\mu\nu} 
(\partial_\mu b)(\partial_\nu b) + \ldots \, .
\label{graviph}
\ee
The appropriate identification is then $\Im U = b$.  Before
integration over $y$, the structure of the $D=4$ kinetic terms in the
$y$-dependent $D=5$ bosonic Lagrangian is identical to the case of a
compactification on $S^1/Z_2$ without warping.

To derive the K\"ahler potential and superpotential of the effective
$D=4$ supergravity, however, we should integrate over $y$
eqs.~(\ref{5Defflag}) and (\ref{graviph}), and compare the result with
the bosonic Lagrangian of $N=1$, $D=4$ Poincar\'e supergravity coupled
to a chiral supermultiplet $U$ in an arbitrary frame:
\bea
\label{sugra4}
{\cal L}_4 & = & - {1\over2} \widehat{e}_4 \Phi_4 
\widehat{R}_4 + {3 \over 4} \widehat{e}_4 \Phi_4 \widehat{g}^{\mu\nu}
(\partial_\mu \log \Phi_4) (\partial_\nu \log \Phi_4)
\nn \\ & &  
- \widehat{e}_4 \Phi_4 (K_{\ov{U}U})^{-1} \widehat{g}^{\mu\nu}
(\partial_\mu \ov{U}) (\partial_\nu U) - \Phi_4^2 \, \widehat{V}_4 \, ,
\eea
where $\widehat{V}_4$ is now the scalar potential in units of the
$D=4$ field-dependent Planck mass $M_4^2[r(x)]=\Phi_4[r(x)]$. If
desired, the dilaton prefactor $\Phi_4$ can be eliminated by a
field-dependent rescaling of the $D=4$ metric $\widehat{g}_{\mu\nu}
(x)$.  However, we do not need to integrate eq.~(\ref{5Defflag})
exactly. Both the original $D=5$ supergravity and the effective $D=4$
supergravity are being considered for small values of their respective
cosmological constants, $\lambda$ and $\lambda_4$, with respect to
their respective Planck masses, $M_5$ and $M_4$. We will then consider
in the following two possible expansions, either in $\lambda_4$ or in
$\lambda \pi r$, and derive the effective $D=4$ supergravity in these
two limits.
\bc 
{\bf 5.1: The \boldmath{$\lambda_4$} expansion} 
\ec
To perform an expansion in $\lambda_4$, we assume that $\lambda_4/\lambda
\ll 1$ and take
\be
\label{asoltext}
a [r(x),y] = \alpha \, e^{\dd - \lambda r |y|}
+ \beta \,  e^{\dd + \lambda r |y|} \, ,
\ee
choosing for simplicity
\be
\label{conv}
\beta \simeq 1 - {\lambda_4^2 \over 4 \, \lambda^2} \, ,
\qquad
\alpha \simeq  {\lambda_4^2 \over 4 \, \lambda^2} \, ,
\ee
which is consistent with the exact expressions for $\alpha$ and $\beta$,
given in eq.~(\ref{solab4}) of the appendix, at leading order in the
expansion parameter. So doing, we assume to be close to the fine-tuned
case $\lambda_0 = \lambda_\pi = - \lambda$. Since the fine-tuned case,
corresponding to $\lambda_4=0$, has vanishing superpotential, the
superpotential of the detuned case must be ${\cal O} ( \lambda_4
)$. Therefore, in a consistent leading-order approximation, we must
evaluate the K\"ahler potential in the limit $\lambda_4 = 0$, and the
scalar potential $\widehat{V}_4$ up to ${\cal O} (\lambda_4^2)$. 

Setting $\alpha=0$ and $\beta=1$, the $D=4$ dilaton prefactor reads:
\be
\label{phi4red}
\Phi_4(r) = {1 \over \lambda} \left( 
e^{\dd 2 \lambda \pi r} - 1 \right) \, .
\ee
With the complexification $U(x) = r(x) + i \, b(x)$ as before, it
corresponds to the K\"ahler potential of eqs.~(\ref{Kbr}) and
(\ref{XUdef}), with a plus sign in the present conventions.  This
result was found in \cite{rsefftun}: the proof that the integrated
axion kinetic term is compatible with the K\"ahler potential
(\ref{Kbr}) is considerably more subtle.

We can also take the potential $\widehat{V}_4(r)$ from
eq.~(\ref{v4fin}) of the appendix, and expand it up to order
$\lambda_4^2/\lambda^2$. Doing so, we obtain the result of
eq.~(\ref{potx2}), with the plus sign and:
\be
\rho_0 = \lambda_4 \, \left( 1 -  e^{\dd -
2 \, \lambda \, \pi \, r_c } \right)^{1/2} \, ,
\qquad
\rho_3 = \lambda_4 \, \left( e^{\dd 
2 \, \lambda \, \pi \, r_c } - 1 \right)^{1/2} \, .
\ee
This potential is minimized, as expected, for
\be
e^{\dd 2 \, \lambda \, \pi \, r} = {\rho_0^2 \over \rho_3^2}
= 
{(\lambda+\lambda_0) \, (\lambda-\lambda_\pi) \over
(\lambda-\lambda_0) \, (\lambda+\lambda_\pi) } \, ,
\ee
and its value at the minimum is, as expected,
\be
\langle \widehat{V}_4 \rangle = - {3 \, \rho_0^2 \, \rho_3^2 
\over \rho_0^2 - \rho_3^2 } = - 3 \, \lambda_4^2 \, .
\ee
We then derive, knowing that there is no potential for the axion
$b(x)$, and using the results of section~4, the effective
superpotential of eq.~(\ref{sup2x}), in agreement with the results
of \cite{rseffdetun}. In the $T$ field basis, this superpotential reads
\be
w_{RS} (T) ={1 \over 2 \sqrt{2}} \left[ \rho_0 \left(
1 - T \right)^3 + \rho_3 \, e^{\dd i \, \varphi} \left( 
1 + T \right)^3 \right] \, ,
\label{wrst}
\ee
and corresponds to one of the generalized $T$ gaugings discussed in
section~3.
\bc 
{\bf 5.2: The \boldmath{$\lambda \pi r$} expansion} 
\ec
If we take the expressions for $\Phi_4(r_c)$ and $\widehat{V}_4(r_c)$
given by eqs.~(\ref{Phi4is})--(\ref{int4}) of the appendix, perform
everywhere the replacement (\ref{replac}), and expand all the
exponentials up to ${\cal O} (\lambda \, \pi \, r)$, we obtain the
following results. The $D=4$ dilaton prefactor,
\be
\Phi_4 = (\alpha + \beta)^2 \, 2 \, \pi \, r(x) \, ,
\ee
gives a K\"ahler potential of the form (\ref{KahlerT}), where we
should now call $T \equiv r(x) + i \, b(x)$ the complex field that was
previously called $U$.

As for the scalar potential $\widehat{V}_4$, if we take $\alpha$ and
$\beta$ to be both of order one we get, at the first non trivial order
in the expansion parameter:
\be
\label{potexp2}
\widehat{V}_4 = {6 \, \lambda^2 \, \alpha \, \beta \over \pi \,
(\alpha + \beta)^2} \, {r_c - 2 \, r(x) \over r^2(x)} \, .
\ee
This is of the general form of eq.~(\ref{pota}), thus we know from
section~3 that it is generated by a superpotential of the form of
eq.~(\ref{supa}). As expected, the potential in eq.~(\ref{potexp2}) is
minimized for
\be
\langle r(x) \rangle = r_c \, ,
\ee
and at the minimum
\be
\langle \widehat{V}_4 \rangle = {- 3 \, \lambda_4^2 \over 2 \,
\pi \, r_c \, (\alpha + \beta)^2 } = - {  3 \, \lambda_4^2 \over
M_4^2} \, .
\ee
As already discussed in section~3, we have a stable $AdS_4$
background, with broken or unbroken supersymmetry according to the
choice of $\langle b(x) \rangle$ along its flat direction.
\vspace{1.3cm}
\bc
{\bf Acknowledgments}
\ec
This work was partially supported by the European Programmes
HPRN-CT-2000-00148 (Across the Energy Frontier) and HPRN-CT-2000-00131
(Quantum Spacetime). The non-resident authors thank the Laboratoire de
Physique Th\'eorique of the Ecole Normale Sup\'erieure and the
Institut de Physique of the Universit\'e de Neuch\^atel for their warm
hospitality and support during part of this project. One of us (F.Z.)
would like to thank R.~Rattazzi for a useful discussion.
\vspace{1.3cm}
%
%
\bc 
{\bf Appendix: backgrounds of the detuned SUSY-RS model} 
\ec
We collect in this appendix some results on the backgrounds \cite{rs1,
kaloper} of the detuned \cite{kaloper, susyrsdetun} SUSY-RS model
\cite{rs1, susyrstun} that can be useful to understand the derivation
of its effective theory \cite{rsefftun, rseffdetun} as presented in
the text. For the purposes of the present paper it is sufficient to
consider only the bosonic gravity sector of the theory, neglecting the
graviphoton and the fermions.

The relevant part of the $D=5$ Lagrangian reads, in units where the
$D=5$ Planck mass $M_5$ is set equal to one:
\be
\cL_g = 
- {1 \over 2} \, e_5 \, R_5 - e_5 \, \Lambda_5 - 6 \, e_4 \, \left[
\lambda_0 \, \delta(y) - \lambda_\pi \, \delta(y - \pi) \right] \, .
\label{lagbos}
\ee
In our notation, $E_M^{\; A}$ is the f\"unfbein, $M=[(\mu=0,1,2,3),5]$
are curved space-time indices, $y = x^5$, $A=[(a = \hat{0}, \hat{1},
\hat{2}, \hat{3}), \hat{5}]$ are flat tangent-space indices, $e_5 =
\det E_M^{\; A}$, $e_4 = \det E_\mu^{\; a}$, $\Lambda_5 \equiv - 6 \,
\lambda^2 < 0$ is the $D=5$ cosmological constant in units of $M_5$,
$R_5$ is the $D=5$ scalar curvature in the conventions of
\cite{wbook}, the $D=5$ Minkowski metric is $\eta_{AB} = diag
(-1,+1,+1,+1,+1)$, and the delta functions are normalized according to
$\int_{- \pi}^\pi \! dy \, \delta (y) = 1$. On the right-hand side of
eq.~(\ref{lagbos}) we understand an overall factor of $M_5^3$, so that
the input parameters $(\lambda, \lambda_0, \lambda_\pi)$ have all the
dimension of a mass. The extrema of integration over $y$ are indeed
$\pm \pi/M_5$, if we want the coordinate $y$ to have the dimension of
a length. Supersymmetry requires $\lambda_0^2, \, \lambda_\pi^2 \le
\lambda^2$: in the following, it will not be restrictive to remove a
twofold ambiguity and assume that $\lambda > 0$ and $- \lambda \le
\lambda_\pi \le \lambda_0 \le + \lambda$.

We are interested in backgrounds of the form 
\be
\label{interval}
ds^2 = a^2 (r_c,y) \, \hat{g}_{\mu \nu} (x) 
dx^\mu dx^\nu + r_c^2 \, dy^2 \, ,
\ee
where $\hat{g}_{\mu \nu}$ is a maximally symmetric $D=4$ metric such
that
\be
\widehat{R}_{\mu \nu} (\widehat{g})  =
- \Lambda_4 \, \widehat{g}_{\mu \nu} \equiv
3 \, \lambda_4^2 \, \widehat{g}_{\mu \nu} \, ,
\ee
so that $\Lambda_4 = - 3 \, \lambda_4^2$ can be interpreted as the
$D=4$ cosmological constant in units of the $D=4$ Planck mass $M_4$,
and the constant $r_c > 0$ can be interpreted, if we identify $y$ with
$y + 2 \pi$, as the compactification radius in units of $M_5$. Notice
that the parameterization of eq.~(\ref{interval}) is redundant, since
we can rescale $a^2(r_c,y)$ by an arbitrary constant factor and
$\widehat{g}_{\mu \nu}(x)$ by its inverse: to fix this ambiguity we
may require, for example, that $a^2(r_c,0)=1$, so that the $D=4$
metric $\widehat{g}_{\mu \nu} (x)$ is identified with the $D=5$ metric
on the brane at $y=0$. However, we can also leave the normalization
factor in $a(r_c,y)$ undetermined, since it can always be reabsorbed
in the definition of the $D=4$ metric and of the $D=4$ Planck mass: it
will be fixed only when we want to give a definite $D=5$ geometrical
interpretation to the $D=4$ metric. As appropriate for the detuned
case \cite{kaloper}, we make the ansatz:
\be
\label{asol}
a (r_c,y) = \alpha \, e^{\dd - \lambda r_c |y|}
+ \beta \,  e^{\dd + \lambda r_c |y|} \, ,
\ee
where $\alpha$ and $\beta$ are real dimensionless constants, and we
understand periodicity for the function $|y|$. The normalization condition
$a^2(r_c,0)=1$ translates into
\be
\label{normal}
\alpha + \beta = \pm 1 \, .
\ee
Depending on the problem under consideration, it may be convenient to
make use of eq.~(\ref{normal}) or to leave the overall normalization
factor in $\alpha$ and $\beta$ undetermined. The bulk equations of
motion require that
\be
\label{lfour}
\lambda_4^2 = 4 \lambda^2 \, \alpha \, \beta \, .
\ee
The equations of motion at the fixed points (in the `upstairs'
picture) require
\be
\label{lzero}
{\lambda_0 \over \lambda} = {\alpha - \beta \over \alpha + \beta} \, ,
\ee
and
\be
\label{lampi}
{\lambda_\pi \over \lambda} = {\alpha \, e^{\dd - \lambda r_c \pi}
- \beta \, e^{\dd + \lambda r_c \pi} \over \alpha \, e^{\dd - 
\lambda r_c \pi} + \beta \, e^{\dd + \lambda r_c \pi}} \, .
\ee
The last two conditions can be fulfilled for
\be
\beta (\lambda + \lambda_0) = \alpha (\lambda - \lambda_0) \, ,
\label{abzero}
\ee
and 
\be
e^{\dd \, 2 \lambda r_c \pi} \,
(\lambda-\lambda_0) \, (\lambda+\lambda_\pi)
=
(\lambda+\lambda_0) \, (\lambda-\lambda_\pi) \, .
\label{fixr}
\ee
Therefore, we can solve $D=5$ Einstein's equations everywhere,
including the fixed points, with constant radius $r_c > 0$: in the
fully detuned case, $- \lambda < \lambda_\pi < \lambda_0 < \lambda$,
we have an $AdS_4$ background ($\lambda_4^2 > 0$), and the radius
$r_c$ is uniquely determined by eq.~(\ref{fixr}); in the fully tuned
case, $\lambda_0 = \lambda_\pi = \pm \lambda$, we have a $Minkowski_4$
($\lambda_4^2 = 0$) background, and $r_c$ is undetermined. Partially
tuned choices of $\lambda_0$ and $\lambda_\pi$ compatible with
supersymmetry lead either to $r_c=0$ ($\lambda_0 = \lambda_\pi \ne \pm
\lambda$) or to $r_c = + \infty$ ($\lambda_0 = \lambda$ and/or
$\lambda_\pi = - \lambda$). Notice that in the tuned case the
background $a(r_c,y)$ is a single exponential, whereas in the fully
detuned case it is always a double exponential. If we also assume the
normalization condition in eq.~(\ref{normal}), we can go further and
find explicitly:
\be
\label{solab4}
\alpha = \pm {\lambda + \lambda_0 \over 2 \, \lambda} \, ,
\qquad
\beta = \pm {\lambda - \lambda_0 \over 2 \, \lambda} \, ,
\qquad
\lambda_4^2 = \lambda^2 - \lambda_0^2 \, .
\ee

With the line element of eq.~(\ref{interval}), the $D=5$ bosonic
Lagrangian of eq.~(\ref{lagbos}) reads
\be
\cL_g =  
- {1 \over 2} \Phi \widehat{e}_4 \widehat{R}_4
+ 6 \, \widehat{e}_4 \,
\left[ { (a')^2 \, a^2 \over r_c}
+ \lambda^2 \, r_c \, a^4
- \lambda_0 \, a^4 \, \delta(y)
+ \lambda_\pi \, a^4 \, \delta(y - \pi) 
\right] \, ,
\label{5Deffpot}
\ee
where $\widehat{e}_4 = |\det \widehat{g}_{\mu\nu}|^{1/2}$,
$\widehat{R}_4$ is the 4D curvature scalar for the metric
$\widehat{g}_{\mu\nu}$, and we have safely neglected total
derivatives. Notice that the $D=4$ Einstein term is not canonically
normalized, but has a $D=5$ dilaton prefactor
\be
\label{bigphi}
\Phi (r_c, \, y) = r_c \, a^2 (r_c, \, y)  \, .
\ee
The rest of eq.~(\ref{5Deffpot}) contains $y$-dependent contributions
to the scalar potential. After we integrate over $y$, the Einstein
term remains non-canonical, with a $D=4$ dilaton prefactor
\be
\label{Phi4is}
\Phi_4(r_c) \equiv \int_{-\pi}^\pi dy \, \Phi (r_c, y) = 
{1 \over \lambda} \left[ \alpha^2 \left( 1 - e^{- 2 \lambda 
\pi r_c} \right) + \beta^2 \left( e^{2 \lambda \pi r_c} - 
1 \right) + 4 \alpha \beta \lambda \pi r_c \right] \, ,
\ee
which can be eliminated by a rescaling of the $D=4$ metric
$\widehat{g}_{\mu \nu}$ or absorbed in the definition of the $D=4$
$r_c$-dependent Planck mass, $M_4^2(r_c) = \Phi_4(r_c)$. Recalling
that we deal we a non-canonical $D=4$ Einstein term, we can now
compute the $D=4$ potential for the background field $r_c$, by
integrating over $y$ the part of eq.~(\ref{5Deffpot}) within square
brackets. It is convenient to normalize the potential in units of
$M_4(r_c)$. If we do so, we obtain:
\be
\label{v4fin}
\widehat{V}_4 (r_c) \equiv {V_4 (r_c) \over M_4^4(r_c)} =  
- 6 \, \Phi_4^{-2} \, \widehat{e}_4 \, \left[ I_{22} (r_c) 
+ I_4 (r_c) - \lambda_0 a^4(r_c,0) + \lambda_\pi a^4(r_c,\pi) 
\right] \, ,
\ee
where 
\be
\label{i22}
I_{22} (r_c) \equiv \int_{-\pi}^\pi \!\!\! dy \, 
{(a')^2 \, a^2 \over r_c}  = 
{\lambda \over 2} \left[ \alpha^4 \left( 1 - e^{- 4 \lambda 
\pi r_c} \right) + \beta^4 \left( e^{4 \lambda \pi r_c} - 
1 \right) - 8 \alpha^2 \beta^2 \lambda \pi r_c \right] \, ,
\ee
and
\bea
I_4 (r_c) \equiv \int_{-\pi}^\pi \!\!\! dy \, a^4 \, 
\lambda^2 \, r_c &  = & {\lambda \over 2} \left[ 
\alpha^4 ( 1 - e^{- 4 \lambda \pi r_c})
+
8 \alpha^3 \beta  ( 1 - e^{- 2 \lambda \pi r_c} ) \right.
\nn \\ & + & 
\left. 
8 \alpha \beta^3  ( e^{2 \lambda \pi r_c} - 1 )
+
\beta^4 (e^{4 \lambda \pi r_c} - 1)
+
24 \alpha^2 \beta^2 \lambda \pi r_c \right] \, .
\label{int4}
\eea
The minimization of $\widehat{V}_4 (r_c)$ is tedious, but of course
reproduces all the results for $r_c$ and $\lambda_4$ obtained from the
$D=5$ equations of motion, as functions of the input parameters
$(\lambda, \lambda_0, \lambda_\pi)$.
\newpage
%
%

%

\begin{thebibliography}{99}
%

\bibitem{effn1}
E.~Witten, Phys.\ Lett.\ B {\bf 155} (1985) 151; \\
J.~P.~Derendinger, L.~E.~Ib\'a\~nez and H.~P.~Nilles,
Nucl.\ Phys.\ B {\bf 267} (1986) 365; \\
S.~Ferrara, C.~Kounnas and M.~Porrati,
Phys.\ Lett.\ B {\bf 181} (1986) 263.

\bibitem{cfgvp}
E.~Cremmer, S.~Ferrara, L.~Girardello and A.~Van Proeyen,
Phys.\ Lett.\ B {\bf 116} (1982) 231 and
Nucl.\ Phys.\ B {\bf 212} (1983) 413.

\bibitem{pure5d}
E.~Cremmer, in {\em Superspace and supergravity}, S.~W.~Hawking and
M.~Rocek eds., Cambridge University Press, 1981, pp. 267-282; \\
A.~H.~Chamseddine and H.~Nicolai,
Phys.\ Lett.\ B {\bf 96} (1980) 89; \\
R.~D'Auria, E.~Maina, T.~Regge and P.~Fre,
Annals Phys.\  {\bf 135} (1981) 237; \\
M.~Gunaydin, G.~Sierra and P.~K.~Townsend,
Nucl.\ Phys.\ B {\bf 242} (1984) 244.

\bibitem{gauged5d}
M.~Gunaydin, G.~Sierra and P.~K.~Townsend,
Nucl.\ Phys.\ B {\bf 253} (1985) 573.

\bibitem{bfzjhep}
J.~Bagger, F.~Feruglio and F.~Zwirner, JHEP {\bf 0202} (2002) 010
[arXiv:hep-th/0108010].

\bibitem{rsefftun}
M.~A.~Luty and R.~Sundrum,
Phys.\ Rev.\ D {\bf 64} (2001) 065012
[arXiv:hep-th/0012158]; \\
J.~Bagger, D.~Nemeschansky and R.~J.~Zhang,
JHEP {\bf 0108} (2001) 057 [arXiv:hep-th/0012163].

\bibitem{rseffdetun}
J.~Bagger and M.~Redi, arXiv:hep-th/0310086 and
arXiv:hep-th/0312220.

\bibitem{fluxes}
S.~B.~Giddings, S.~Kachru and J.~Polchinski,
Phys.\ Rev.\ D {\bf 66} (2002) 106006
[arXiv:hep-th/0105097]; \\
S.~Kachru, M.~B.~Schulz and S.~Trivedi, JHEP {\bf 0310} (2003) 007
[arXiv:hep-th/0201028]; \\
R.~D'Auria, S.~Ferrara and S.~Vaula, 
New J.\ Phys.\ {\bf 4} (2002) 71 
[arXiv:hep-th/0206241]; \\
S.~Ferrara and M.~Porrati, Phys.\ Lett.\ B {\bf 545} (2002) 411
[arXiv:hep-th/0207135]; \\
S.~Kachru, M.~B.~Schulz, P.~K.~Tripathy and S.~P.~Trivedi,
JHEP {\bf 0303} (2003) 061 [arXiv:hep-th/0211182]; \\
L.~Andrianopoli, R.~D'Auria, S.~Ferrara and M.~A.~Lledo, 
JHEP {\bf 0303} (2003) 044 
[arXiv:hep-th/0302174]; \\
C.~Angelantonj, S.~Ferrara and M.~Trigiante,
JHEP {\bf 0310} (2003) 015
[arXiv:hep-th/0306185];\\
P.~K.~Tripathy and S.~P.~Trivedi, 
JHEP {\bf 0303} (2003) 028 
[arXiv:hep-th/0301139]; \\
A.~Giryavets, S.~Kachru, P.~K.~Tripathy and S.~P.~Trivedi,
arXiv:hep-th/0312104.

\bibitem{dwvp}
B.~de Wit and A.~Van Proeyen,
Nucl.\ Phys.\ B {\bf 245} (1984) 89.

\bibitem{seven}
E.~Cremmer, C.~Kounnas, A.~Van Proeyen, J.~P.~Derendinger, 
S.~Ferrara, B.~de
Wit and L.~Girardello, Nucl.\ Phys.\ B {\bf 250} (1985) 385.

\bibitem{dfmp}
J.~P.~Derendinger, S.~Ferrara, A.~Masiero and A.~Van Proeyen,
Phys.\ Lett.\ B {\bf 140} (1984) 307.

\bibitem{N=4}
M.~de Roo,
Nucl.\ Phys.\ B {\bf 255} (1985) 515 and 
Phys.\ Lett.\ B {\bf 156} (1985) 331;\\
E.~Bergshoeff, I.~G.~Koh and E.~Sezgin, 
Phys.\ Lett.\ B {\bf 155} (1985) 71; \\
M.~de Roo and P.~Wagemans, 
Nucl.\ Phys.\ B {\bf 262} (1985) 644.

\bibitem{adk}
I.~Antoniadis, J.~P.~Derendinger and C.~Kounnas, Nucl.\ Phys.\ B 
{\bf 551} (1999) 41 [arXiv:hep-th/9902032].

\bibitem{GivP}
A.~Giveon and M.~Porrati,
Phys.\ Lett.\ B {\bf 246} (1990) 54 and
Nucl.\ Phys.\ B {\bf 355} (1991) 422.

\bibitem{ak}
I.~Antoniadis and C.~Kounnas, 
Phys.\ Lett.\ B {\bf 261} (1991) 369. 

\bibitem{kk}
E.~Kiritsis and C.~Kounnas, 
Ê
Nucl.\ Phys.\ B {\bf 503} (1997) 117 
[arXiv:hep-th/9703059]. 

\bibitem{hetflux}
J.~P.~Derendinger, L.~E.~Ib\'a\~nez and H.~P.~Nilles,
Phys.\ Lett.\ B {\bf 155} (1985) 65; \\
M.~Dine, R.~Rohm, N.~Seiberg and E.~Witten,
Phys.\ Lett.\ B {\bf 156} (1985) 55.

\bibitem{fklz}
S.~Ferrara, C.~Kounnas, D.~L\"ust and F.~Zwirner, 
Ê
Nucl.\ Phys.\ B {\bf 365} (1991) 431. 

\bibitem{fkpz}
S.~Ferrara, C.~Kounnas, M.~Porrati and F.~Zwirner,
Nucl.\ Phys.\ B {\bf 318} (1989) 75.

\bibitem{PZ}
M.~Porrati and F.~Zwirner, 
Nucl.\ Phys.\ B {\bf 326} (1989) 162. 

\bibitem{scsc}
J.~Scherk and J.~H.~Schwarz,
Phys.\ Lett.\ B {\bf 82} (1979) 60
and Nucl.\ Phys.\ B {\bf 153} (1979) 61.

\bibitem{rs1}
L.~Randall and R.~Sundrum,
Phys.\ Rev.\ Lett.\  {\bf 83} (1999) 3370
[arXiv:hep-ph/9905221].

\bibitem{susyrstun}
R.~Altendorfer, J.~Bagger and D.~Nemeschansky,
Phys.\ Rev.\ D {\bf 63} (2001) 125025
[arXiv:hep-th/0003117]; \\
T.~Gherghetta and A.~Pomarol,
Nucl.\ Phys.\ B {\bf 586} (2000) 141
[arXiv:hep-ph/0003129]; \\
A.~Falkowski, Z.~Lalak and S.~Pokorski,
Phys.\ Lett.\ B {\bf 491} (2000) 172
[arXiv:hep-th/0004093] and
Nucl.\ Phys.\ B {\bf 613} (2001) 189
[arXiv:hep-th/0102145]; \\
E.~Bergshoeff, R.~Kallosh and A.~Van Proeyen,
JHEP {\bf 0010} (2000) 033
[arXiv:hep-th/0007044].

\bibitem{kaloper}
N.~Kaloper, Phys.\ Rev.\ D {\bf 60} (1999) 123506
[arXiv:hep-th/9905210].

\bibitem{susyrsdetun}
P.~Brax, A.~Falkowski and Z.~Lalak, 
Phys.\ Lett.\ B {\bf 521} (2001) 105 
[arXiv:hep-th/0107257];  \\
J.~Bagger and D.~V.~Belyaev,
Phys.\ Rev.\ D {\bf 67} (2003) 025004
[arXiv:hep-th/0206024] and
JHEP {\bf 0306} (2003) 013
[arXiv:hep-th/0306063]; \\
Z.~Lalak and R.~Matyszkiewicz, Nucl.\ Phys.\ B {\bf 649} (2003) 389
[arXiv:hep-th/0210053],
Phys.\ Lett.\ B {\bf 562} (2003) 347 [arXiv:hep-th/0303227]
and arXiv:hep-th/0310269; \\
P.~Brax and N.~Chatillon, 
JHEP {\bf 0312} (2003) 026
[arXiv:hep-th/0309117].

\bibitem{noscale}
E.~Cremmer, S.~Ferrara, C.~Kounnas and D.~V.~Nanopoulos,
Phys.\ Lett.\ B {\bf 133} (1983) 61; \\
J.~R.~Ellis, C.~Kounnas and D.~V.~Nanopoulos,
Nucl.\ Phys.\ B {\bf 241} (1984) 406 and
Nucl.\ Phys.\ B {\bf 247} (1984) 373; \\
J.~R.~Ellis, A.~B.~Lahanas, D.~V.~Nanopoulos and K.~Tamvakis,
Phys.\ Lett.\ B {\bf 134} (1984) 429.

\bibitem{stability}
P.~Breitenlohner and D.~Z.~Freedman, 
Phys.\ Lett.\ B {\bf 115} (1982) 197 and
Annals Phys.\  {\bf 144} (1982) 249; \\
P.~K.~Townsend, Phys.\ Lett.\ B {\bf 148} (1984) 55.

\bibitem{wbook}
S.~Weinberg, {\em Gravitation and Cosmology}, John Wiley and Sons,
1972.

\end{thebibliography}
\end{document}